\documentclass[conference]{IEEEtran}
\IEEEoverridecommandlockouts

\usepackage[utf8]{inputenc}
\usepackage{amsmath,amssymb,amsfonts}
\usepackage{graphicx}
\usepackage[dvipsnames]{xcolor}
\usepackage{microtype}
\usepackage{dblfloatfix}
\usepackage{flushend}
\usepackage{xcolor}
\usepackage{booktabs}
\usepackage{tabularx}
\usepackage{multirow}
\usepackage{colortbl}
\usepackage{array}
\usepackage[flushleft]{threeparttable}
\usepackage{makecell}
\usepackage[export]{adjustbox}
\usepackage{enumitem}

\usepackage{tikz}
\usetikzlibrary{arrows.meta, positioning, fit, backgrounds, calc}

\usepackage[caption=false,font=footnotesize]{subfig}

\usepackage[numbers, sort&compress]{natbib}
\usepackage{hyperref}
\hypersetup{colorlinks=true,linkcolor=blue,citecolor=blue,urlcolor=magenta}

\usepackage{orcidlink}
\usepackage{pifont}
\newcommand{\cmark}{\textcolor{green!55!black}{\ding{51}}}
\newcommand{\xmark}{\textcolor{red!60!black}{\ding{55}}}
\newcommand{\pmark}{\textcolor{orange!80!black}{\textbf{\textasciitilde}}}

\newcommand{\totalrecords}{$\sim$1{,}890{,}000\ }
\newcommand{\totalrecordscompact}{1.89M\ }
\newcommand{\perclassrecords}{270{,}000\ }
\newcommand{\trainingrecords}{189{,}000\ }

\def\BibTeX{{\rm B\kern-.05em{\sc i\kern-.025em b}\kern-.08em
    T\kern-.1667em\lower.7ex\hbox{E}\kern-.125emX}}

\definecolor{simblue}{RGB}{210,228,255}
\definecolor{atkred}{RGB}{255,218,218}
\definecolor{featorange}{RGB}{255,237,210}
\definecolor{datateal}{RGB}{210,245,240}
\definecolor{idspurple}{RGB}{235,220,255}
\definecolor{rowgray}{gray}{0.93}
\definecolor{rowhighlight}{RGB}{230,245,255}
\definecolor{roadgray}{RGB}{180,180,180}
\definecolor{pavegray}{RGB}{220,220,220}
\definecolor{vehblue}{RGB}{60,100,200}
\definecolor{siggreen}{RGB}{50,180,80}

\begin{document}

\title{SPADE: SPaT Attack Detection from the \\ Connected Vehicle's Perspective}

\author{
    \IEEEauthorblockN{
        James Di Novo$^1$\IEEEauthorrefmark{2}\,\orcidlink{0009-0008-4910-8617},~%
        \IEEEmembership{Graduate Student Member,~IEEE},~%
        Hany Ragab$^2$\IEEEauthorrefmark{2}\,\orcidlink{0000-0003-3167-9100},~%
        \IEEEmembership{Member,~IEEE},\\%
        Sylvain P. Leblanc$^3$\IEEEauthorrefmark{2}\,\orcidlink{0000-0001-7882-1229},~%
        \IEEEmembership{Senior Member,~IEEE}
    }

\thanks{\IEEEauthorrefmark{2}Department of Electrical and Computer Engineering, Faculty of Engineering, Royal Military College of Canada (RMC), Kingston, Ontario, K7K~7B4, Canada. Emails: \href{mailto:james.di-novo@rmc-cmr.ca}{\color{black}{james.di-novo@rmc-cmr.ca}}$^{1}$, hany.ragab@\{\href{mailto:hany.ragab@rmc-cmr.ca}{\color{black}{rmc-cmr}}, \href{mailto:hany.ragab@queensu.ca}{\color{black}{queensu}}\}.ca$^{2}$, \href{mailto:sylvain.leblanc@rmc.ca}{\color{black}{sylvain.leblanc@rmc.ca}}$^{3}$.}

}

\maketitle

\begin{abstract}
Signal Phase and Timing (SPaT) messages are a cornerstone of connected
vehicle (CV) safety, enabling CVs to perceive and respond to intersection
state through Vehicle-to-Infrastructure (V2I) and Vehicle-to-Vehicle (V2V)
communication. The integrity of these messages is threatened by a range of
application-layer attacks that can bypass conventional authentication when a
roadside unit or peer vehicle is compromised. Existing intrusion detection
research either defends the infrastructure side or targets V2V Basic Safety
Message (BSM) / Cooperative Awareness Message (CAM) misbehavior, leaving the onboard CV perspective on SPaT integrity unaddressed. To close this gap, we introduce SPADE --- the \textbf{SP}aT \textbf{A}ttack \textbf{D}etection
and \textbf{E}valuation dataset --- a labelled, multi-modal, simulation-based
dataset designed specifically for deep learning IDS research in this space.
SPADE is generated through Eclipse MOSAIC using runtime attack injection at
the SAE J2735 application layer across six attack classes and one benign
class. By combining four intersection geometries, six operating conditions,
and five independent random-seed repetitions, SPADE comprises 180 unique base
scenario runs, yielding \totalrecords labelled timestep records (\perclassrecords per class). Each record fuses SPaT message fields, onboard camera confidence
scores, and cooperative V2V peer data across 40 features, reflecting the
multi-modal signal space required to distinguish deliberate attacks from
environmental degradation. The dataset, generation code, and scenario
configurations are released publicly to support reproducible and comparative
IDS research in C-V2X security. The developed toolbox, instructions, and dataset link are publicly available on GitHub: \texttt{\url{https://github.com/jdinovo/SPADE}}.
\end{abstract}


\section{Introduction}
\label{sec:intro}

Intelligent Transportation Systems (ITS) rely on Road Side Units (RSUs)
embedded in Traffic Signal Control Systems (TSCS) to broadcast Signal Phase
and Timing (SPaT) messages over Dedicated Short-Range Communication (DSRC)
or Cellular Vehicle-To-Everything (C-V2X) channels~\cite{abdelhakeem2025}.
These messages give Connected Vehicles (CVs) machine-readable awareness of
intersection state, enabling on-the-fly signal timing optimization and safe
traversal in conditions where visual signal perception is
unreliable~\cite{feng2022}. As C-V2X deployments mature under the SAE J2735
standard~\cite{j2735_202409_2024}, SPaT becomes an increasingly critical input to automated driving
functions, making its integrity a safety imperative.

\begin{figure}[htbp]
	\centerline{\includegraphics[scale=0.7]{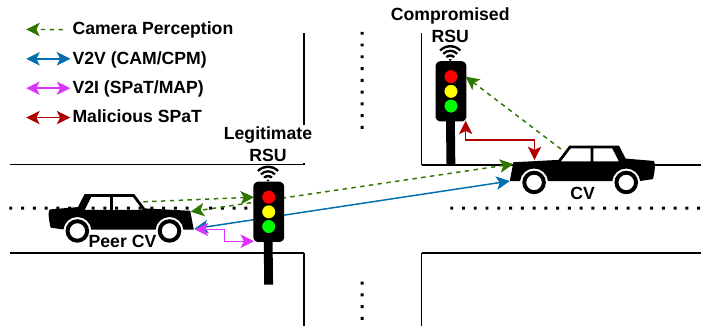}}
    \caption{Overview of the SPADE dataset focal point.}
    \label{fig:spat_overview}
\end{figure}

The same connectivity that makes SPaT valuable also makes is an attack
surface. A compromised RSU or peer CV can inject false phase states, replay
stale messages, manipulate countdown timers, flood competing identities,
suppress broadcasts, or impersonate a legitimate
RSU~\cite{abdelhakeem2025}. In each case, the receiving CV has no way to
detect the attack through cryptographic means alone, since valid credentials
are assumed to be available on the compromised node. An IDS running onboard
the CV and operating at the application message layer is therefore a
necessary complement to the underlying public key infrastructure. Fig.~\ref{fig:spat_overview} exemplifies a simplified scenario where a CV is receiving malicious SPaT data from a compromised TSCS which conflicts with perception and peer data.

Despite the recognized severity of these threats, the IDS literature has not
produced a purpose-built, multi-modal dataset for this specific problem.
Existing work addresses either traffic signal intrusion detection from the
infrastructure perspective~\cite{feng2022, chowdhury2023, shen2023, almallah2023}
or V2V BSM misbehavior detection~\cite{van2018veremi, veremi2026}. To the best of our knowledge supported by a recent comprehensive survey, there exists no dataset that captures the onboard CV perspective of SPaT data integrity combined with onboard vehicle sensors and V2V communications~\cite{abdelhakeem2025}.

SPADE addresses this gap directly. The dataset is generated in Eclipse
MOSAIC~\cite{mosaic2022} through runtime attack injection at the SAE J2735 application layer,
ensuring the multi-modal data streams --- SPaT/MAP, camera perception, and CAM/Cooperative Perception Messages (CPM) --- are
internally consistent and reflect what an actual CV would observe under each
attack condition. With 180 unique base scenario runs, each executed once per
label class, SPADE comprises \totalrecords records with \perclassrecords per
class --- a scale comparable to recent deep learning (DL) IDS work on vehicular
datasets~\cite{vemisnet2025}. The main contributions are:
\begin{enumerate}[label=\arabic*), leftmargin=*, noitemsep]
    \item A publicly released labelled dataset of \totalrecords multi-modal
          records covering six SPaT attack classes and a benign class,
          generated across 180 unique scenario configurations in Eclipse
          MOSAIC.
    \item A documented threat model mapping each attack type to its targeted
          SAE J2735 fields and observable multi-modal signatures.
    \item A feature set combining SPaT message fields, camera confidence, and
          V2V cooperative data designed to support classical and deep
          learning IDS approaches, with a scenario-level train/validate/test separation
          to prevent data leakage.
    \item An extensible scenario configuration covering four intersection
          geometries, six operating conditions, and five random-seed
          repetitions, enabling controlled intra-class diversity.
\end{enumerate}

The remainder of this paper details related work in Section \ref{sec:related}, defines the threat model in Section \ref{sec:threat}, details the dataset generation and characteristics in Sections \ref{sec:generation} and \ref{sec:description}, discusses future work in Section \ref{sec:future_work}, and concludes in Section \ref{sec:conclusion}.
\section{Related Work and Dataset Gap}
\label{sec:related}

\subsection{Traffic Signal Security}

Feng et al.~\cite{feng2022} provide a comprehensive treatment of TSCS
cybersecurity with CVs, modeling attack scenarios in which adversarial
vehicles feed false data to the infrastructure controller to manipulate
signal timing. Chowdhury et al.~\cite{chowdhury2023} adopt an
evidence-theoretic approach to traffic signal IDS, also from the
infrastructure perspective, while Shen et al.~\cite{shen2023} use
infrastructure-side sensors and traffic invariants to detect data spoofing
in connected vehicle-based signal control. Al Mallah et
al.~\cite{almallah2023} target resilience at the adaptive multi-agent
traffic controller level. In all four cases, detection is performed by
infrastructure components, not by the receiving CV.

The General Motors patent (US~11{,}875{,}677~B2)~\cite{gm_patent2024}
represents the prior work closest to what we propose, outlining a method for CVs to identify
misbehavior in V2I communications. It does not produce a labelled dataset,
does not employ machine learning, or incorporate cross-modal data. Abdel Hakeem and Kim~\cite{abdelhakeem2025} survey ML, federated learning, and edge AI for V2X IDS broadly, confirming that ML-based onboard detection of SPaT-specific attacks remains open and that no existing dataset addresses it.

\subsection{V2X and Automotive Datasets}

The VeReMi~\cite{van2018veremi} and VeReMi NextGen~\cite{veremi2026} datasets
are the best-established labelled datasets for V2X misbehavior detection.
Both target V2V BSM spoofing in VANETs across 180 scenario configurations
spanning fifteen attack types, four scenarios, and three dataset splits; neither addresses SPaT messages or V2I attacks, and neither
incorporates camera or cross-modal fusion data. The ROAD
dataset~\cite{verma_comprehensive_2024} provides actual vehicular attack data including
replay, jamming, and message falsification scenarios, but it is collected from
the infrastructure perspective and does not include SPaT-specific attack
classes. The Car-Hacking Dataset~\cite{carhacking2018} and
CICIoV2024~\cite{cicioV2024} cover intra-vehicle CAN bus attacks, an
entirely different attack surface from V2X communications. Venkatasamy et
al.~\cite{venkatasamy2024} produce a labelled V2X IDS dataset using
cryptographic protocol features, but target general VANET communications
rather than SPaT-specific attack signatures at the CV receiver. Recent Deep-Learning work on the VeReMi Extension evaluates LSTM and BiLSTM architectures at
dataset scales from 100K to 2M samples and demonstrates that \perclassrecords samples
per class produces competitive F1 scores for 7-class
classification~\cite{vemisnet2025}, motivating SPADE's per-class target.

\subsection{Simulation of Urban Mobility (SUMO) Scenarios}

The Ingolstadt Traffic Scenario (InTAS) provides a detailed and accurate representation of Ingolstadt, Germany~\cite{lobo_intas_2020}. Lobo et al.~\cite{lobo_intas_2020} have meticulously ensured that each street and intersection adequately represents the topology of the city's roads. Traffic includes public transport and vulnerable road users. InTAS provides an invaluable resource; however, there are significant differences between North American and European road networks and traffic. 
McKenney et al.~\cite{mckenney_distributed_2013} created a realistic traffic signal scenario, published in 2013, based on a section of downtown Ottawa, Canada. This scenario features 50 traffic signals spanning a variety of road types from low-volume, single lane, residential streets to multi-lane, high-volume, main arterial roadways~\cite{mckenney_distributed_2013}. Due to the nature of the primary objective this scenario was restricted to a relatively small section of downtown Ottawa which limits the amount of data which could be generated in order to train and test large models. The scenario was also created before SPaT was fully implemented.

\subsection{Gap Summary}

Table~\ref{tab:dataset_comparison} summarizes the gap across the most
relevant prior work and datasets. SPADE is the first labelled dataset that
combines CV-side perspective, SPaT specificity, onboard cross-modal fusion, and
deep-learning-ready labelling at scale across seven balanced classes. SPADE will utilize a novel Simulation of Urban Mobility (SUMO)~\cite{lopez_microscopic_2018} scenario generated for the city of Ottawa, Canada. Ottawa presents a wide variety of North American road types with numerous complex signaled intersections making it an ideal city for a SPaT-centric dataset. 

The six selected attack types, as explained in section \ref{sec:threat}, are specifically applicable to SPaT messages. These attacks have been examined in other areas of research, such as against CAM and BSM, as well as TSCS from the perspective of the TSCS. They have yet to be applied to SPaT messages from the perspective of the CV. 

\begin{table*}[t]
\centering
\vspace{-2pt}
\caption{Comparison of related datasets and works.
         \cmark~= fully addressed, \pmark~= partially, \\ \xmark~= not
         addressed. ``Infra'' = infrastructure-side detection.}
\label{tab:dataset_comparison}
\renewcommand{\arraystretch}{1.25}
\begin{threeparttable}
\footnotesize
\begin{tabular}{l c c c c c c c}
\toprule
\multirow{2}{*}{\textbf{Work / Dataset}} &
\multirow{2}{*}{\textbf{Year}} &
\multirow{2}{*}{\textbf{Perspective}} &
\multirow{2}{*}{\makecell{\textbf{SPaT-}\\\textbf{Specific}}} &
\multirow{2}{*}{\makecell{\textbf{Cross-Modal}\\\textbf{Fusion}}} &
\multirow{2}{*}{\makecell{\textbf{Labelled}\\\textbf{Dataset}}} &
\multirow{2}{*}{\makecell{\textbf{DL-}\\\textbf{Ready}}} &
\multirow{2}{*}{\makecell{\textbf{Attack}\\\textbf{Classes}}} \\
& & & & & & & \\
\midrule
\rowcolor{rowgray}
Feng et al.~\cite{feng2022}               & 2022 & Infra.  & \cmark & \xmark & \xmark & \xmark & 3  \\
Chowdhury et al.~\cite{chowdhury2023}     & 2023 & Infra.  & \cmark & \xmark & \xmark & \xmark & 2  \\
\rowcolor{rowgray}
Shen et al.~\cite{shen2023}               & 2023 & Infra.  & \cmark & \pmark & \xmark & \xmark & 1  \\
Al Mallah et al.~\cite{almallah2023}      & 2023 & Infra.  & \xmark & \xmark & \xmark & \xmark & -- \\
\rowcolor{rowgray}
GM Patent~\cite{gm_patent2024}            & 2024 & CV     & \cmark & \xmark & \xmark & \xmark & -- \\
VeReMi~\cite{van2018veremi}                   & 2018 & CV     & \xmark & \xmark & \cmark & \cmark & 15  \\
\rowcolor{rowgray}
VeReMi NextGen~\cite{veremi2026}          & 2026 & CV     & \xmark & \xmark & \cmark & \cmark & 15  \\
ROAD~\cite{verma_comprehensive_2024}                      & 2024 & Vehic. & \xmark & \xmark & \cmark & \cmark & 4  \\
\rowcolor{rowgray}
Car-Hacking~\cite{carhacking2018}         & 2018 & CAN    & \xmark & \xmark & \cmark & \cmark & 4  \\
CICIoV2024~\cite{cicioV2024}              & 2024 & CAN    & \xmark & \xmark & \cmark & \cmark & 5  \\
\rowcolor{rowgray}
Venkatasamy et al.~\cite{venkatasamy2024} & 2024 & CV     & \xmark & \xmark & \cmark & \cmark & 5  \\
Lobo et al.~\cite{lobo_intas_2020} & 2020 & Infra.    & \xmark & \xmark & \xmark & \xmark & -  \\
\midrule
\rowcolor{rowhighlight}
\textbf{SPADE (Ours)} & \textbf{2026} & \textbf{CV} &
    \cmark & \cmark & \cmark & \cmark & \textbf{6} \\
\bottomrule
\end{tabular}
\begin{tablenotes}
\item \pmark~Shen et al.\ use infrastructure-side sensors alongside traffic
      invariants, not onboard CV cross-modal fusion. ROAD covers real-world
      vehicular attacks but does not address SPaT specifically.
\end{tablenotes}
\end{threeparttable}
\end{table*}

\section{Threat Model}
\label{sec:threat}

\subsection{Assumptions}

The threat model targets the SAE J2735 application message layer as seen by
a receiving CV, consistent with the V2X attack taxonomy
in Abdel Hakeem and Kim~\cite{abdelhakeem2025}. Three assumptions bound the scope of our work.

\textit{Attacker node.} The attacker controls either a compromised RSU or a
compromised peer CV holding valid PKI credentials (IEEE 1609.2
certificates~\cite{ieee_ieee_2025}), so message authentication does not flag the transmissions.
Detection must rely on message content and cross-modal consistency.

\textit{Attacker goal.} The attacker seeks to cause the CV to misinterpret
intersection state, whether by inducing incorrect phase perception,
triggering a premature or delayed stop, creating conflicting information
across sources, or suppressing legitimate SPaT broadcasts.

\textit{Scope boundary.} Physical-layer jamming, GPS spoofing, CAN bus
attacks, and adversarial ML attacks on trained models are excluded. All six
attack types are implemented as modifications to the SPaT message at the
application layer, keeping the feature space coherent and avoiding a
dependency on network-layer simulation. 

We are focusing on the application layer from the CV's perspective. There are numerous attack types that apply to other perspectives, such as the TSCS, or other network layers; however, the IDS we are considering will be applied at the application layer. The IDS will be reading unencrypted SPaT, MAP, CAM, and CPM transmitted from other CVs and TSCS. It will also have perception, proprioceptive, exteroceptive data gathered from inter-vehicle sensors. From the perspective of the CV, the TSCS could be seen as compromised from the onset of its interaction, or become compromised partially through the interaction.

\subsection{Attack Classes}

Table~\ref{tab:threat_model} defines the six attack classes and the benign
class which form the SPADE label schema. These are relevant attacks that would affect SPaT messages or CPMs received by a CV. Denial of Service (DoS) is modeled
from the receiving CV's perspective as message suppression rather than
channel flooding, which would require network-layer simulation.
Suppression is the application-layer observable effect of DoS and is
detectable through the inter-message interval feature without requiring
ns-3 or OMNeT++ integration.

\begin{table}[t]
\centering
\caption{SPADE threat model: attack classes, targeted SAE J2735
         fields, and observable multi-modal signatures.}
\label{tab:threat_model}
\renewcommand{\arraystretch}{1.3}
\begin{threeparttable}
\footnotesize
\begin{tabularx}{\linewidth}{
    >{\raggedright\arraybackslash}p{0.24\linewidth}
    >{\raggedright\arraybackslash}p{0.35\linewidth}
    >{\raggedright\arraybackslash}X
}
\toprule
\textbf{Class} & \textbf{J2735 Field(s)} & \textbf{Observable Signature} \\
\midrule
\rowcolor{rowgray}
\texttt{BENIGN}         & N/A & All streams consistent; camera-SPaT agreement holds \\
\texttt{FALSE\_STATE}   & \texttt{MovementPhaseState} & Reported phase contradicts camera detection \\
\rowcolor{rowgray}
\texttt{REPLAY}         & Full SPaT msg & Stale phase/timer; timestamp inconsistent with clock \\
\texttt{TIMING\_MANIP}  & \texttt{MinEndTime}, \texttt{MaxEndTime} & Phase matches camera; countdown deviates from elapsed time \\
\rowcolor{rowgray}
\texttt{SYBIL}          & Source ID, V2V msg & Peer consensus fails; conflicting phase perception reports \\
\texttt{DOS}            & N/A (suppression)\pmark & Inter-message interval exceeds expected broadcast period \\
\rowcolor{rowgray}
\texttt{IMPERSONATION}  & \texttt{IntersectionID},\newline source address & Source mismatch with known RSU identity \\
\bottomrule
\end{tabularx}
\begin{tablenotes}
\item \pmark DoS is modeled as application-layer message suppression, observable
      as the absence of SPaT broadcasts within the expected reception window.
\end{tablenotes}
\end{threeparttable}
\end{table}

\section{SPADE Dataset Generation}
\label{sec:generation}

\subsection{Simulation Environment}

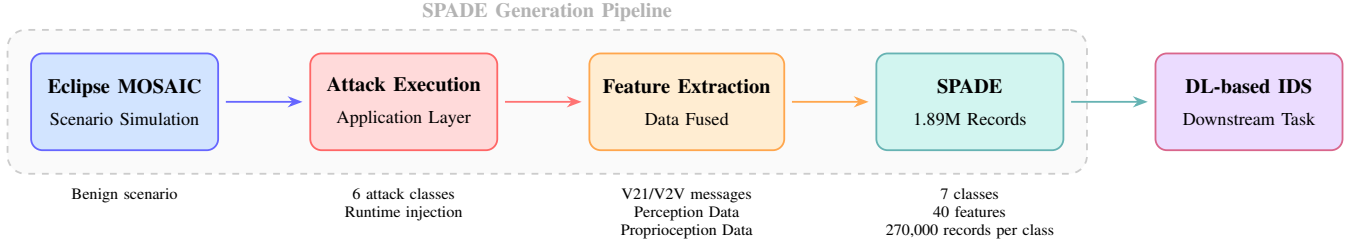
\begin{figure*}[t]
\vspace{-10pt}
\centering
\resizebox{0.98\linewidth}{!}{%
\begin{tikzpicture}[
    font=\small, >=Stealth,
    simnode/.style={draw=blue!60,fill=simblue,rounded corners=5pt,
        minimum width=2.9cm,minimum height=1.5cm,align=center,inner sep=6pt,thick},
    atknode/.style={draw=red!60,fill=atkred,rounded corners=5pt,
        minimum width=2.9cm,minimum height=1.5cm,align=center,inner sep=6pt,thick},
    featnode/.style={draw=orange!70,fill=featorange,rounded corners=5pt,
        minimum width=2.9cm,minimum height=1.5cm,align=center,inner sep=6pt,thick},
    datanode/.style={draw=teal!60,fill=datateal,rounded corners=5pt,
        minimum width=2.9cm,minimum height=1.5cm,align=center,inner sep=6pt,thick},
    idsnode/.style={draw=purple!60,fill=idspurple,rounded corners=5pt,
        minimum width=2.9cm,minimum height=1.5cm,align=center,inner sep=6pt,thick},
    sublbl/.style={font=\scriptsize,align=center,text width=2.9cm},
    arr/.style={->,thick,shorten >=3pt,shorten <=3pt}
]
\node[simnode] (sim) {\textbf{Eclipse MOSAIC}\\[4pt]\footnotesize Scenario Simulation};
\node[atknode,right=1.4cm of sim]  (atk)  {\textbf{Attack Execution}\\[4pt]\footnotesize Application Layer};
\node[featnode,right=1.4cm of atk] (feat) {\textbf{Feature Extraction}\\[4pt]\footnotesize Data Fused};
\node[datanode,right=1.4cm of feat](data) {\textbf{SPADE}\\[4pt]\footnotesize \totalrecordscompact Records};
\node[idsnode,right=1.4cm of data] (ids)  {\textbf{DL-based IDS}\\[4pt]\footnotesize Downstream Task};
\draw[arr,blue!60]   (sim)--(atk);
\draw[arr,red!55]    (atk)--(feat);
\draw[arr,orange!70] (feat)--(data);
\draw[arr,teal!60]   (data)--(ids);
\node[sublbl,below=0.45cm of sim]  {Benign scenario};
\node[sublbl,below=0.45cm of atk]  {6 attack classes\\Runtime injection};
\node[sublbl,below=0.45cm of feat] {V21/V2V messages\\Perception Data\\Proprioception Data};
\node[sublbl,below=0.45cm of data] {7 classes\\40 features\\\perclassrecords records per class};
\node[sublbl,below=0.45cm of ids]  {};
\begin{scope}[on background layer]
  \node[draw=gray!40,dashed,rounded corners=8pt,thick,
        fit=(sim)(atk)(feat)(data),inner sep=10pt,fill=gray!4,
        label={[font=\small\bfseries,gray!60]above:SPADE Generation Pipeline}]{};
\end{scope}
\end{tikzpicture}}
\vspace{-6pt}
\caption{SPADE generation and usage pipeline. Attacks are
         injected at runtime, ensuring all separate data streams reflect a consistent scenario.}
\label{fig:pipeline}
\end{figure*}

SPADE is generated in Eclipse MOSAIC~\cite{mosaic2022}, an open
co-simulation framework integrating traffic mobility, vehicle
applications, and V2X network simulation. MOSAIC's Application Simulator
enables custom V2X applications to run on individual simulation
objects, making it the appropriate layer for attack injection at the SAE J2735
level. Base intersection geometries are exported from OpenStreetMap (OSM)~\cite{noauthor_openstreetmap_nodate},
ensuring road layouts reflect real-world configurations. Every CV in the 
scenario is an SAE level 4 and beyond autonomous vehicle which does not require human intervention. This futuristic approach, where all vehicles are autonomously 
driven with no human behind the wheel, means that all decisions are made based on 
what is perceived by sensors and received through the V2X network. This prevents vehicle decisions from being interrupted by a human, enabling predictable and consistent outcomes. Vehicles transmit Cooperative Awareness Messages (CAMs), receive full SPaT messages from the TSCS, and log SPaT content, V2V peer data, and onboard camera perception to per-vehicle CSV files. Fig.~\ref{fig:pipeline} illustrates the complete generation pipeline.

The perception of signals and vehicles is done through MOSAICs built-in methods, which do not make use of any confidence levels in the current version. Perceptual occlusion is taken into account, however only as far as full obstruction by objects such as other vehicles or buildings~\cite{mosaic2022}. Occlusion due to weather or lens distortion is not a factor. 

Environmental degradation per scenario is captured by a visibility score $V_s \in [0,1]$, where $V_s = 0.95$ denotes clear conditions and $V_s = 0.20$ represents severe impairment like dense fog. Separately, camera confidence $C_{\mathrm{cam}}$ is modeled geometrically from the distance to the traffic light $d$ (meters), angular offset $\alpha$ (degrees), maximum range $R$ (meters), and field of view $A$ (degrees) as:
\begin{align}
r_d      &= \min\!\left(\tfrac{d}{R}, 1\right),                         \\
r_\alpha &= \min\!\left(\tfrac{|\alpha|}{A/2}, 1\right),                \\
F_d      &= \max\left(0,\; 1 - r_d^2\right),                            \\
F_\alpha &= \max\left(0,\; 1 - r_\alpha^2\right),                       \\
C_{\mathrm{base}} &= 0.4 + 0.6 \cdot F_d \cdot F_\alpha,                \\
C_{\mathrm{cam}}  &= \operatorname{clamp}(C_{\mathrm{base}}
    + \varepsilon,\; 0,\; 1), \quad \varepsilon \sim \mathcal{N}(0,\;0.03^2),
\end{align}

where $\operatorname{clamp}(x,0,1) = \max(0,\min(1,x))$. The quadratic
factors $F_d$ and $F_\alpha$ decay detection quality with distance and
off-axis angle; the floor of $0.4$ models residual detectability at the
geometric boundary of the camera's field of view. Gaussian noise
$\varepsilon$ introduces per-sample variability and the result is bounded
to $[0,1]$. The observed phase and $C_{\mathrm{cam}}$ are broadcast to
surrounding vehicles via CPM over the V2V channel.

\subsection{Scenario Configurations}

\begin{figure*}[t]
\centering
\subfloat[\texttt{2-LANE}\label{fig:s_2lane}]{\includegraphics[width=0.15\linewidth, valign=c]{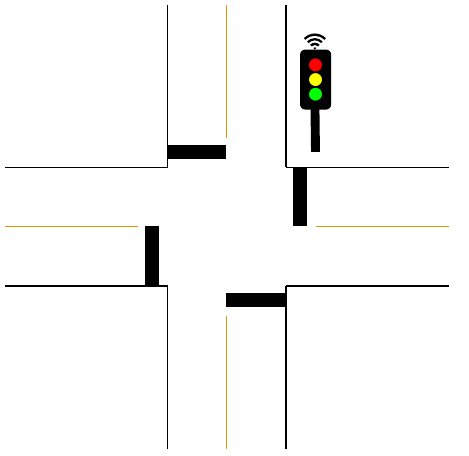}}
\hfill
\subfloat[\texttt{4-LANE}\label{fig:s_4lane}]{\includegraphics[width=0.15\linewidth, valign=c]{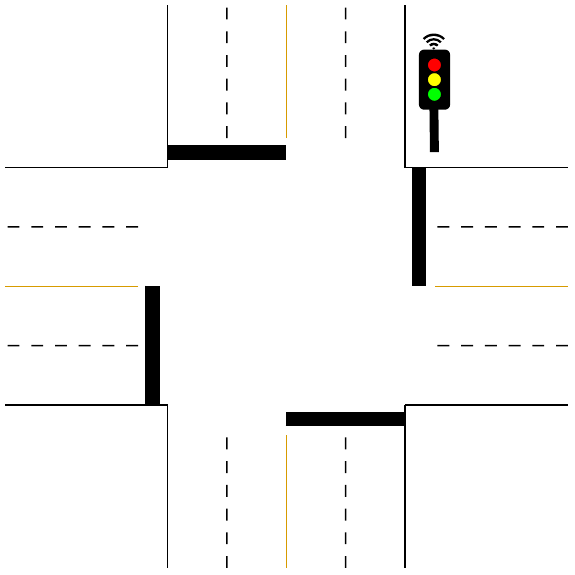}}
\hfill
\subfloat[\texttt{T-CROSS}\label{fig:s_tcros}]{\includegraphics[width=0.15\linewidth, valign=c]{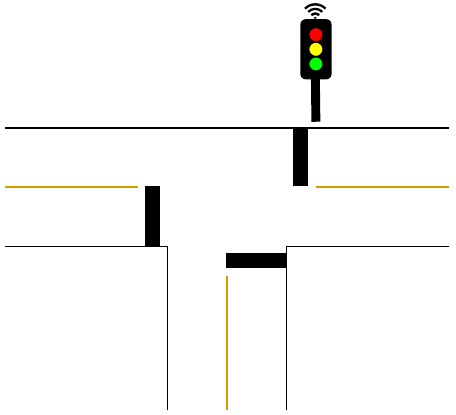}}
\hfill
\subfloat[\texttt{PXO}\label{fig:s_pxo}]{\includegraphics[width=0.15\linewidth, valign=c]{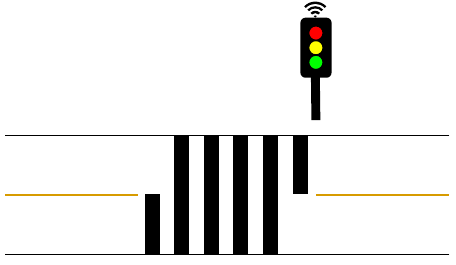}}
\\[5pt]
\subfloat[\texttt{LOW-DEN}\label{fig:s_lowden}]{\includegraphics[width=0.15\linewidth, valign=c]{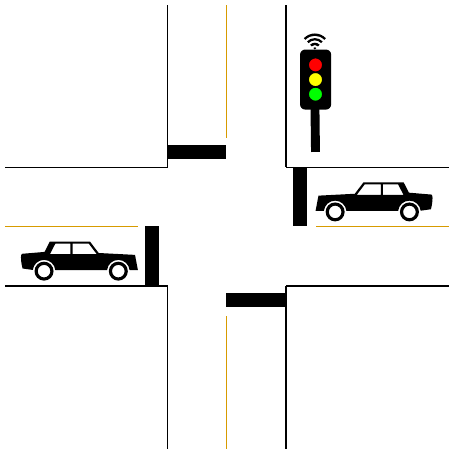}}
\hfill
\subfloat[\texttt{HIGH-DEN}\label{fig:s_highden}]{\includegraphics[width=0.15\linewidth, valign=c]{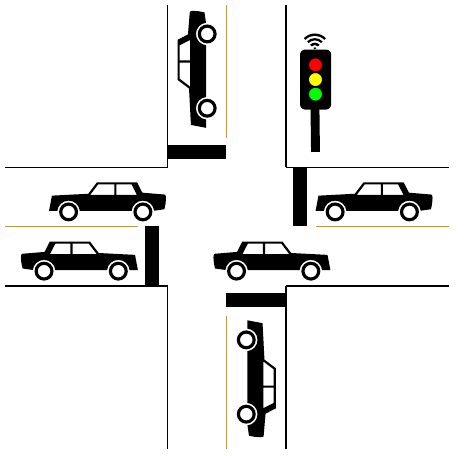}}
\hfill
\subfloat[\texttt{HIGH-VIS}\label{fig:s_highvis}]{\includegraphics[width=0.15\linewidth, valign=c]{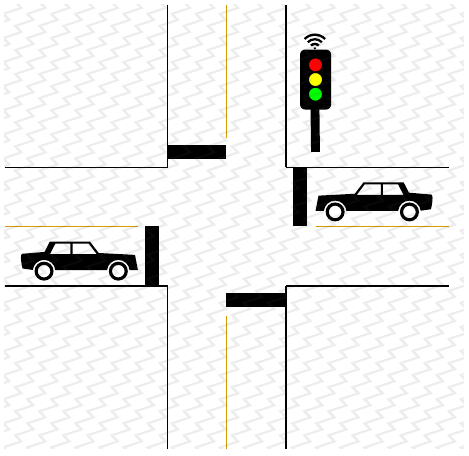}}
\hfill
\subfloat[\texttt{LOW-VIS}\label{fig:s_lowvis}]{\includegraphics[width=0.15\linewidth, valign=c]{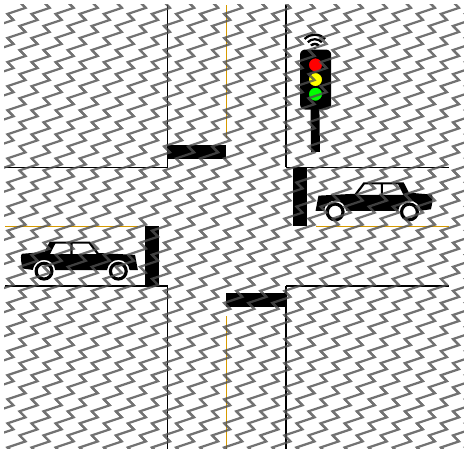}}
\caption{The eight simulation scenario configuration types shown for
         SPADE. Note that there is a traffic signal for every stop line, however only one is shown in each configuration example. Row~1 (\ref{fig:s_2lane}--\ref{fig:s_pxo}): intersection
         and mid-block crossing geometry types derived from OSM exports. Row~2 (\ref{fig:s_lowden}--\ref{fig:s_lowvis}):
         representative extremes of the two operating condition dimensions
         --- traffic density (\texttt{LOW-DEN} and \texttt{HIGH-DEN}) and camera visibility (clear \texttt{HIGH-VIS} and heavily degraded \texttt{LOW-VIS}). The full configuration matrix also includes
         \texttt{MED-DEN} and \texttt{MED-VIS} as intermediate levels:
         $4 \times 3 \times 3 = 36$ base configurations, each repeated
         with five random seeds, yielding 180 unique scenario runs.}
\label{fig:scenarios}
\end{figure*}

SPADE is generated across ten scenario configuration types organized along three dimensions, as shown in Fig.~\ref{fig:scenarios}. The four intersection
geometries are a standard two-lane four-way crossing
(Fig.~\ref{fig:s_2lane}), a four-lane four-way crossing
(Fig.~\ref{fig:s_4lane}), a T-shaped intersection (Fig.~\ref{fig:s_tcros}), and a pedestrian-actuated mid-block crossing (PXO) (Fig.~\ref{fig:s_pxo}). The three operating
conditions are low traffic density (Fig.~\ref{fig:s_lowden}), medium density, and high density (Fig.~\ref{fig:s_highden}). There are also visibility modifiers for
high visibility (Fig.~\ref{fig:s_highvis}), medium visibility, or low visibility (Fig.~\ref{fig:s_lowvis}) of the camera representing adverse weather occluding the lens.

The four geometry types and six operating conditions combine into 36 base
scenario configurations. Each base configuration is then repeated with five
independent random seeds to vary vehicle route assignments and inter-arrival
times, yielding \textbf{180 unique scenario runs}. Every run is executed
seven times --- once per label class --- for a total of \textbf{1260
simulation executions}. Each run generates a varied number of records between each unique scenario dependent on the density and proximity of traffic, and the specific attack being executed. Some attacks generate more records, such as replay and false state, while others, such as DOS, generate fewer. Records are recorded from each CVs perspective, so the number of CVs is proportional to the number of records generated for a given scenario. If the density is high and CVs are perceivable between each other, there would be more records generated as a result, due to camera perception and V2V communications. Road layout, vehicle density, and camera
conditions are held constant across the seven label runs of each scenario,
ensuring that differences in the exported data are attributable to the
attack type rather than to scenario variation.

\subsection{Runtime Attack Injection}

Attacks are injected at runtime rather than applied as post-hoc modifications
to exported CSV data. Each attack class is implemented as a custom
V2X application on a malicious RSU or compromised peer CV node in MOSAIC. Simulating attacks within MOSAIC ensures that the multi-modal data sources remain synchronized and accurately reflect the scenario as excepted. The attack classes are as follows, corresponding to relevant classes in Abdel Hakeem and Kim~\cite{abdelhakeem2025}:

\texttt{FALSE\_STATE}: A \texttt{MaliciousRSU} application broadcasts
modified \texttt{MovementPhaseState} values, equivalent to the falsified
information attack category.

\texttt{REPLAY}: A \texttt{ReplayAttacker} captures SPaT messages in a first
benign pass and re-injects them with a configurable delay in a second pass,
consistent with the replay attack definition.

\texttt{TIMING\_MANIP}: The \texttt{MaliciousRSU} leaves
\texttt{MovementPhaseState} correct while modifying \texttt{MinEndTime} and
\texttt{MaxEndTime} in \texttt{TimeChangeDetails}, corresponding to the
timing attack class.

\texttt{SYBIL}: Multiple phantom CV nodes transmit conflicting perceived signal state
over V2V, creating the peer consensus failure described.

\texttt{DOS}: The legitimate RSU application is silenced for a defined
interval, producing an SPaT absence detectable via inter-message
interval.

\texttt{IMPERSONATION}: A rogue node transmits SPaT messages with a spoofed
\texttt{IntersectionID} or source address, matching the impersonation attack
description.

\subsection{Ground-Truth Labelling}

The active attack class at each simulation timestep is logged to a file for easy corroboration with generated data. Ground-truth labels are an inherent byproduct of the
simulation rather than a separate annotation step: labels are assigned by
joining the scenario schedule to the exported CSV on timestamp, requiring no
manual annotation.

\section{Dataset Description}
\label{sec:description}

\subsection{Class Distribution, Split, and Balance Rationale}

SPADE contains \totalrecords records distributed equally across seven classes:
approximately \perclassrecords records per class. Class balance reflects the primary use case of
training deep learning IDS models, for which class imbalance suppresses
minority-class gradient contributions and makes standard accuracy metrics
misleading~\cite{abdelhakeem2025}. Researchers evaluating real-world
deployment conditions should apply class-weighting or calibration strategies
to reflect the benign-heavy distribution of operational SPaT traffic. The number of records generated per class run is correlated to the number of vehicles running in the simulation and the amount of time the simulation runs. As such, lower density classes are run for longer to ensure the number of records generated is kept similar. 

The dataset is partitioned at the \emph{scenario-run} level to prevent
data leakage: records from the same simulation run appear in exactly one
partition. Of the 180 unique base runs, 126 are assigned to training (70\%), 27
to validation (15\%), and 27 to testing (15\%), giving \trainingrecords training,
40{,}500 validation, and 40{,}500 test records per class. This is
fundamentally different from random record-level splitting, which would place
windows from the same simulation run in both training and test sets and
inflate reported detection rates.

\subsection{Feature Set}

Records consist of a combined 40 features from six sources: the received SPaT and corresponding MAP message, the onboard camera perception, V2V peer CAM and shared perception message, and vehicle proprioceptive data.
Table~\ref{tab:feature_set} describes the full feature set. The inclusion of both onboard and shared perception data enable the opportunity for a more robust IDS which does not solely rely on changes in SPaT data alone to determine legitimacy. Perceptive data can be referenced to identify discrepancies. 

\begin{table}[p]
\centering
\vspace{-10pt}
\caption{SPADE feature set: SPADE Data Records (40)}
\label{tab:feature_set}
\renewcommand{\arraystretch}{1.2}
\begin{threeparttable}
\footnotesize
\begin{tabularx}{\linewidth}{
    >{\raggedright\arraybackslash}p{0.34\linewidth}
    >{\raggedright\arraybackslash}X
}
\toprule
\textbf{Feature}  & \textbf{Description} \\
\midrule
    \multicolumn{2}{l}{Vehicle Data Record (11)} \\
\midrule
\rowcolor{rowgray}
\texttt{vehicle\_id} & ID of the current vehicle \\
\texttt{route\_id} & Route which the vehicle is traveling \\
\rowcolor{rowgray}
\texttt{lane\_id} & Current lane index of the vehicle \\
\texttt{is\_parked} & Is the vehicle currently in a parked state \\
\rowcolor{rowgray}
\texttt{position} & Latitude and longitude of the vehicle \\
\texttt{heading} & Vehicle's current heading \\
\rowcolor{rowgray}
\texttt{speed} & Current vehicle speed \\
\texttt{throttle} & Amount of throttle applied \\
\rowcolor{rowgray}
\texttt{brake} & Amount of brake applied \\
\texttt{longitudinal\_accel} & Amount of longitudinal acceleration \\
\rowcolor{rowgray}
\texttt{slope} & Current slope the vehicle is on \\
\midrule
    \multicolumn{2}{l}{CAM Data Record (5)} \\
\midrule
\rowcolor{rowgray}
\texttt{vehicle\_id} & ID of the transmitting vehicle \\
\texttt{message\_time} & Time the message was sent \\
\rowcolor{rowgray}
\texttt{position} & Current transmitting vehicle position \\
\texttt{distance} & Distance between vehicles \\
\rowcolor{rowgray}
\texttt{vehicle\_awareness} & Transmitting vehicle proprioceptive data points \\
\midrule
    \multicolumn{2}{l}{TL Perception Data Record (10)} \\
\midrule
\rowcolor{rowgray}
\texttt{signal\_id} & ID of the perceived traffic signal \\
\texttt{incoming\_lane} & Lane which precedes the signal \\
\rowcolor{rowgray}
\texttt{distance} & Distance from current vehicle to signal \\
\texttt{angle\_off\_center} & Angle off center of camera \\
\rowcolor{rowgray}
\texttt{signal\_red} & Red one hot \\
\texttt{signal\_yellow} & Yellow one hot \\
\rowcolor{rowgray}
\texttt{signal\_green} & Green one hot \\
\texttt{camera\_conf} & Perception confidence level \\
\rowcolor{rowgray}
\texttt{cur\_vis\_score} & Current visibility score \\
\texttt{conf\_vis\_weighted\textsuperscript{1}} & Confidence level weighted with visibility \\
\midrule
    \multicolumn{2}{l}{Shared TL Perception Data Record (3)} \\
\midrule
\rowcolor{rowgray}
\texttt{vehicle\_id} & ID of the transmitting vehicle \\
\texttt{vehicle\_pos} & Position of the transmitting vehicle \\
\rowcolor{rowgray}
\texttt{signal\_perceptions} & List of perceived signals (TL Percep. records) \\
\midrule
    MAP Data Record (6) & \\
\midrule
\rowcolor{rowgray}
\texttt{message\_time} & Time the message was sent \\
\texttt{intersection\_id} & ID of the intersection \\
\rowcolor{rowgray}
\texttt{lane\_id} & The intersection lane being referenced \\
\texttt{lane\_attributes} & Lane properties (direction, type, shared) \\
\rowcolor{rowgray}
\texttt{lane\_connections} & Connected lanes (ex. u-turn, left, straight) \\
\texttt{position} & Latitude and longitude of the intersection \\
\midrule
    SPaT Data Record (5) & \\
\midrule
\rowcolor{rowgray}
\texttt{signal\_id} & ID of this individual signal \\
\texttt{message\_time} & Time the message was sent \\
\rowcolor{rowgray}
\texttt{signal\_state} & Current signal state \\
\texttt{max\_end\_time} & Maximum length of the current state \\
\rowcolor{rowgray}
\texttt{min\_end\_time} & Minimum length of the current state \\
\bottomrule
\end{tabularx}
\begin{tablenotes}
\item $\textsuperscript{1}\texttt{conf\_vis\_weighted} = \texttt{camera\_conf} \times \texttt{cur\_vis\_score}$
\end{tablenotes}
\end{threeparttable}
\end{table}

\subsection{Dataset Scale Justification}

The \perclassrecords records per class target is motivated by empirical findings
in vehicular DL IDS research. Recent work evaluating LSTM and BiLSTM
architectures on the VeReMi Extension dataset across scales from 100K to
2M samples demonstrates that \perclassrecords samples per class falls within the range
where competitive macro-F1 scores are achieved for multi-class vehicular
IDS, and that performance gains above this scale are progressively
diminishing~\cite{vemisnet2025}. The 40-feature input space is sufficiently
compact that generalization risk is dominated by scenario diversity rather
than raw sample count, which is why SPADE's 180 unique base runs are the
primary design lever. 

\section{Future Work}
\label{sec:future_work}

SPADE is released as a dataset contribution; DL model training and comparative benchmarking represent ongoing work being tackled by the authors.

\subsection{Deep Learning}

Based on the survey of DL methods for V2X IDS by Abdel Hakeem and Kim~\cite{abdelhakeem2025},
three architecture families are natural baselines: a one-dimensional
convolutional neural network (1D-CNN); a stacked LSTM network, which was shown effective by Youness et al.~\cite{vemisnet2025}; and transformer-based encoder with self-attention over the full input window, which has been successfully applied to other CV IDS such as Li et al.~\cite{li_attentionguard_2025}. 

\subsection{Limitations and Future Directions}

When making use of SPADE for training and validation of an IDS the following should be known. SPADE operates under a closed-world assumption: the seven label classes are
exhaustive within the dataset, and models trained on SPADE may not detect
hybrid or parametrically novel attacks outside the training distribution.
This is a common limitation of simulation-based IDS datasets noted
by Abdel Hakeem and Kim~\cite{abdelhakeem2025} and is inherent to the controlled labelling
methodology. Potential follow-on work can address this through: (i) training and
benchmarking the three baseline architectures on SPADE to establish a
comparative leaderboard; (ii) evaluating detection under class-imbalanced
conditions reflecting realistic benign-heavy deployment; (iii) extending
scenarios to include adversarial ML attack variants targeting the trained
IDS; and (iv) hardware-in-the-loop validation using physical C-V2X OBU
hardware to assess sim-to-real transfer. 

The SPaT data generated is on a fixed timing schedule, there are no dynamic SPaT timings which are adjusted based on traffic volume or other factors. The incorporation of such advanced SPaT data would create more difficult scenarios for an IDS to detect malicious messages. 

Due to current SUMO limitations vehicles are unable to operate on the falsified malicious messages generated from attacks. Allowing them to do so requires also allowing collisions which would create noise in the dataset. To address this, the decision the vehicle would make based on it's information is simply logged. The addition of CARLA~\cite{dosovitskiy_carla_2017} or similar to enable collision avoidance while allowing vehicles to act on incorrect information would be ideal. This limitation renders proprioceptive data less useful for training since it always aligns with ground-truth and will remain indistinguishable between attack and benign scenarios. Proprioceptive data still remains relevant in cases such as Sybil attacks, for potentially identifying vehicles which do not actually exist, by making use of CPMs in combination with claimed position.

\section{Conclusion}
\label{sec:conclusion}

SPADE is the first labelled, multi-modal dataset designed for deep
learning-based intrusion detection of SPaT message attacks from the onboard
perspective of a connected vehicle. It fills a documented gap in the C-V2X
security literature, where existing work either
defends the infrastructure side or targets V2V BSM misbehavior. With
\totalrecords balanced records (\perclassrecords per class) drawn from 180 unique
scenario runs, SPADE reaches a scale that is empirically sufficient for LSTM
and Transformer-based IDS training while maintaining
a scenario diversity comparable to the VeReMi benchmark.
The scenario-run-level split prevents data leakage, and the 40-feature
multi-modal design encodes the cross-stream consistency signals that
distinguish deliberate attacks from environmental degradation. SPADE is
released publicly alongside generation code and
scenario configurations.

\bibliographystyle{IEEEtran}
\bibliography{spade_refs}

\end{document}